\documentclass[reprint,showpacs,preprintnumbers,pre,superscriptaddress]{revtex4-1}
\usepackage{graphicx}
\usepackage{amssymb}
\usepackage{amsmath}
\usepackage{amsfonts,color}
\begin{document}

\title{Shortcuts to isothermality and nonequilibrium work relations}
\author{Geng Li}
\affiliation{Department of Physics, Beijing Normal University, Beijing 100875, China}
\author{H. T. Quan}
\affiliation{School of Physics, Peking University, Beijing 100871, China}
\affiliation{Collaborative Innovation Center of Quantum Matter, Beijing 100871, China}
\author{Z. C. Tu}\email[Corresponding author. Email: ]{tuzc@bnu.edu.cn}
\affiliation{Department of Physics, Beijing Normal University, Beijing 100875, China}

%\date{\today}

\begin{abstract}
In conventional thermodynamics, it is widely acknowledged that the realization of an isothermal process for a system requires a quasi-static controlling protocol. Here we propose and design a strategy to realize a finite-rate isothermal transition from an equilibrium state to another one at the same temperature, which is named shortcut to isothermality. By using shortcuts to isothermality, we derive three nonequilibrium work relations, including an identity between the free energy difference and the mean work due to the potential of the original system, a Jarzynski-like equality, and the inverse relationship between the dissipated work and the total driving time. We numerically test these three relations by considering the motion of a Brownian particle trapped in a harmonic potential and dragged by a time-dependent force.
\pacs{05.70.Ln, 05.40.-a}
\end{abstract}
\maketitle

\section{Introduction}
Recent advances in biological protein motors~\cite{Howard2001,Yildiz2004,Block2007,Toyabe2011,Millic2014} and artificial molecular machines~\cite{Balzani2000,Tu2005,Fletcher2005,Kay2007} have boosted the interest of the scientific community in small systems. In contrast to macroscopic systems, such nanometric objects contain much fewer entities than Avogadro's number, which leads to two significant features of small systems: large fluctuations and being easily driven away from equilibrium. Standard methods of statistical mechanics and thermodynamics are not applicable to the understanding of nonequilibrium phenomena of small systems. The Jarzynski equality~\cite{Jarzynski1997} and fluctuation theorems~\cite{Evans1994,GallavottiPRL1995,Crooks1999,Evans2002,Seifert2005,Liu2009,Esposito2010} are two masterpieces in the field of nonequilibrium processes. They generalize fundamental thermodynamic relations to small systems in nonequilibrium situations. Despite the achievement of these two kinds of relations, our understanding of nonequilibrium thermodynamics in small systems is still limited. It is desirable to discover new identities holding true in nonequilibrium processes of small systems.

Besides, transitions between equilibrium states at the same temperature in finite time are also key issues when we investigate nonequilibrium processes of small systems. In conventional thermodynamics, it is widely acknowledged that the realization of an isothermal process requires a quasi-static controlling protocol. Therefore, it takes an infinitely long time to realize transitions between two equilibrium states if the system always keeps in equilibrium with the thermal reservoir. Naturally, an interesting question is whether we can realize transitions between two equilibrium states at the same temperature in finite time (in between, the system is not necessary in equilibrium with the thermal reservoir). A novel concept of engineered swift equilibration~\cite{Martinez2015,Cunuder2016} throws light on the way to realizations of finite-rate isothermal transitions. Mart\'{i}nez \emph{et al.} designed a protocol of engineered swift equilibration for a Brownian particle in a harmonic potential with time-dependent strength, which allows the system to reach equilibrium 100 times faster than the natural equilibration time~\cite{Martinez2015}. Le Cunuder \emph{et al.} realized the fast equilibrium switch of a micro mechanical oscillator by using engineered swift equilibration~\cite{Cunuder2016}. These works illustrate that engineered swift equilibration is a prospective scheme to implement finite-rate isothermal transitions. However, since the method of engineered swift equilibration is limited to the system of a Brownian particle, a unified theoretical framework in general situations is still lacking. Besides, this method begins with choosing the target distribution which alone is inadequate to determine the Hamiltonian and the free energy simultaneously. Thus it is hard to discuss nonequilibrium work relations in finite-rate isothermal transitions based on the concept of engineered swift equilibration.

In this work, we construct a unified framework to conduct finite-rate isothermal transitions. Within this framework, we subsequently derive three nonequilibrium work relations. An auxiliary potential is introduced to the system of interest which is initially in equilibrium, such that the evolution of the system in contact with a thermal reservoir is enforced in the instantaneous equilibrium distribution corresponding to the original Hamiltonian. The total system Hamiltonian can return to the original system Hamiltonian at the beginning and end of the driving process if we impose proper constraints on the controlling protocol. As a result, the distribution functions at two endpoints of the driving process become equilibrium distributions, which means we have realized an isothermal transition without the requirement of quasi-static control, i.e., we have realized a finite-rate isothermal transition. Such a theoretical framework is named shortcut to isothermality. We design the strategy of shortcuts to isothermality for the system of a Brownian particle in both the overdamped and underdamped situations, respectively. Based on shortcuts to isothermality, we subsequently obtain three key relations for nonequilibrium processes. The first relation, (\ref{Eq-freeenergy2}), indicates that the free energy difference equals the mean work due to the time-dependent potential of the original system. The second relation, (\ref{Eq-general-meanwork2}), implies that the dissipated work is inversely proportional to the driving time. The third relation, (\ref{Eq-third-property}), is a Jarzynski-like equality for arbitrary initial distributions. Finally, we numerically check these three relations by considering the motion of a Brownian particle trapped in a harmonic potential and dragged by a time-dependent force.

\section{Definition of shortcuts to isothermality}
Consider a system in contact with a thermal reservoir with a constant temperature $T$. When the external potential varies with time, the Hamiltonian of the system is denoted by $H_{0}(\Gamma,\lambda(t))$, where $\Gamma$ represents the microstate of the system and $\lambda(t)$ is the controlling parameter. The evolution of the distribution function $\rho(\Gamma,t)$ in phase space is determined by
\begin{equation}  \frac{\partial \rho(\Gamma,t)}{\partial t} = \hat{L}_{0}(\Gamma,\lambda(t))\rho(\Gamma,t),  \label{Eq-evolution-equation}\end{equation}
where $\hat{L}_{0}(\Gamma,\lambda(t))$ represents the evolution operator. We restrict our framework to dynamics with a unique equilibrium state, i.e., for fixed $\lambda$ the system will relax towards a unique equilibrium distribution
$\rho_{\mathrm{eq}} \propto \mathrm{e}^{-\beta H_{0}(\Gamma,\lambda)}$,
where $\beta=1/k_{B}T$ with $k_{B}$ being the Boltzmann's constant.
If we introduce an auxiliary potential $U_{1}(\Gamma,t)$ to the Hamiltonian of the system, the total system Hamiltonian becomes
\begin{equation}H(\Gamma,t)=H_{0}(\Gamma,\lambda(t)) + U_{1}(\Gamma,t).\label{Eq-total-Hamiltonian}\end{equation}
The evolution equation (\ref{Eq-evolution-equation}) is modified to the form:
\begin{equation}  \frac{\partial \rho(\Gamma,t)}{\partial t} = \hat{L}_{0}(\Gamma,\lambda(t))\rho(\Gamma,t) + \hat{L}_{1}(\Gamma,t)\rho(\Gamma,t), \label{Eq-evolution-equation2}\end{equation}
with $\hat{L}_{1}(\Gamma,t)$ representing the operators induced by $U_{1}(\Gamma,t)$. At any time, the auxiliary potential $U_{1}(\Gamma,t)$ can escort the evolution of the system so that the system distribution is always in the instantaneous equilibrium distribution of the original Hamiltonian $H_{0}(\Gamma,\lambda(t))$
\begin{equation}\rho(\Gamma,t)= \rho_{\mathrm{ieq}}(\Gamma,\lambda(t)) \equiv \mathrm{e}^{\beta \left [ F(\lambda(t))- H_{0}(\Gamma,\lambda(t))\right]}.\label{Eq-instant-equilibriumDF}\end{equation}
In the above equation,
\begin{equation}F(\lambda) \equiv -\beta^{-1}\ln \left[\int \mathrm{e}^{-\beta H_{0}(\Gamma,\lambda)}\mathrm{d}\Gamma \right]\label{Eq-freeenergy}\end{equation}
represents the free energy of the original system in equilibrium state when the value of $\lambda$ is given. The integral in Eq.~(\ref{Eq-freeenergy}) is over the whole phase space of $\Gamma$. We impose additional constraints on the auxiliary potential. That is, it vanishes at two endpoints of the driving process. With these requirements on $U_{1}(\Gamma,t)$, the distribution functions at two endpoints of the driving process become equilibrium distributions of $H(\Gamma,t)$, which means we have realized a finite-rate isothermal transition. This strategy is enlightened by the idea of shortcuts to adiabaticity for isolated systems~\cite{Emmanouilidou2000,Demirplak2003,Berry2009,Chen2010,Jarzynski2013,Campo2013,Deffner2014,Tu2014}, so we dub it shortcut to isothermality.

The above framework looks similar to the method of escorted free energy simulations put forward by Vaikuntanathan and Jarzynski~\cite{Vaikuntanathan2008}. By loading a proper artificial flow field on the system of interest and introducing a peculiar definition of work, they could generate trajectories where the work on each trajectory equals the free energy difference of the system. In the present work, by contrast, we aim at constructing a unified theoretical framework for implementing finite-rate isothermal transitions. Here, we introduce an auxiliary potential rather than an artificial flow field to escort the evolution of the system. We can still follow the definition of the trajectory dependent work in stochastic thermodynamics~\cite{Seifert2012,Sekimoto2010} without modifying the definition of work as done by Vaikuntanathan and Jarzynski~\cite{Vaikuntanathan2008}.

\section{Shortcuts to isothermality for a Brownian particle in the overdamped situation}
Consider a Brownian particle in a 1-dimensional time-dependent potential $U_{0}(x,\lambda(t))$, with $\lambda(t)$ representing the controlling parameter. The inertial effect can be neglected in the overdamped situation. We introduce an auxiliary time-dependent potential $U_{1}(x,t)$ to the system and the whole potential of the system can be written as
\begin{equation}U(x,t)=U_{0}(x,\lambda(t)) + U_{1}(x,t).\label{Eq-overdamped-Hamiltonian}\end{equation}
The evolution of the distribution function $\rho(x,t)$ for the Brownian particle is governed by the Fokker-Planck equation~\cite{Gardiner1985}
\begin{equation}\frac{\partial \rho}{\partial t} = \frac{1}{\gamma} \frac{\partial}{\partial x}\left[ \frac{\partial (U_{0}+U_{1})}{\partial x}\rho + \frac{1}{\beta}\frac{\partial \rho}{\partial x} \right],\label{Eq-overdamped-FPequation}\end{equation}
where $\gamma$ is the coefficient of friction. According to our assumption, the system evolves along the instantaneous equilibrium state of $U_{0}(x,\lambda(t))$, so the distribution function satisfies
\begin{equation}\rho(x,t) =\rho_{\mathrm{ieq}}(x,\lambda(t)) \equiv \mathrm{e}^{\beta [F(\lambda(t))-U_{0}(x,\lambda(t))]},\label{Eq-overdamped-equilibrium}\end{equation}
with $F(\lambda) \equiv -\beta^{-1}\ln [\int_{-\infty}^{+\infty} \mathrm{e}^{-\beta U_{0}(x,\lambda)}\mathrm{d}x]$. Substituting Eq.~(\ref{Eq-overdamped-equilibrium}) into the Fokker-Planck equation~(\ref{Eq-overdamped-FPequation}), we obtain the equation for $U_{1}(x,t)$:
\begin{equation} \frac{1}{\gamma \beta} \frac{\partial^{2} U_{1}}{\partial x^{2}} - \frac{1}{\gamma}\frac{\partial U_{0}}{\partial x} \frac{\partial U_{1}}{\partial x}  =\left ( \frac{\mathrm{d} F}{\mathrm{d} \lambda} - \frac{\partial U_{0}}{\partial \lambda} \right ) \dot{\lambda}. \label{Eq-overdamped-U1limitation}\end{equation}
 In this work, the dot above a variable represents the time derivative of that variable. Comparing two sides of Eq.~(\ref{Eq-overdamped-U1limitation}), we find $U_{1}(x,t)$ can be written in the following form
\begin{equation}U_{1}(x,t)= \dot{\lambda}(t)f(x,\lambda(t)).\label{Eq-overdamped-U1form}\end{equation}
Substituting it back into Eq.~(\ref{Eq-overdamped-U1limitation}) and then eliminating $\dot{\lambda}(t)$ on both sides, we obtain an ordinary differential equation for $f(x,\lambda(t))$,
\begin{equation} \frac{1}{\gamma \beta} \frac{\partial^{2} f}{\partial x^{2}} - \frac{1}{\gamma}\frac{\partial U_{0}}{\partial x} \frac{\partial f}{\partial x} =\frac{\mathrm{d} F}{\mathrm{d} \lambda} - \frac{\partial U_{0}}{\partial \lambda}. \label{Eq-overdamped-flimitation}\end{equation}
Note that Eq.~(\ref{Eq-overdamped-equilibrium}) is the integrating factor of the above equation. Considering this fact, we can easily solve Eq.~(\ref{Eq-overdamped-flimitation}) and obtain $\partial f / \partial x$ and then $f(x,\lambda(t))$. Substituting the expression of $f(x,\lambda(t))$ into Eq.~(\ref{Eq-overdamped-U1form}), we obtain an analytic form for the auxiliary potential:
\begin{equation}U_{1}(x,t)= \gamma \beta \dot{\lambda}(t) \int\mathrm{d}x\frac{\int\mathrm{d} x h(x,\lambda(t))}{\rho_{\mathrm{ieq}}(x,\lambda(t))}, \label{Eq-overdamped-U1solution}\end{equation}
where
$h(x,\lambda) \equiv \left ( \frac{\mathrm{d} F(\lambda)}{\mathrm{d} \lambda} - \frac{\partial U_{0}(x,\lambda)}{\partial \lambda} \right ) \rho_{\mathrm{ieq}}(x,\lambda)$.
We would like to mention that the idea (extracting potential from the evolution of the distribution function) has been adapted in the study of engineered swift equilibration (see Eq.~(2) of the supplementary information of Ref.~\cite{Martinez2015}). But our focus is different. We extract the auxiliary potential from the given $U_{0}(x,\lambda(t))$ rather than the given target distribution which alone is inadequate to determine the Hamiltonian and the free energy simultaneously. It is this difference that enables us to continue discussing nonequilibrium work relations.

To ensure that  $U(x,t)$ identifies $U_{0}(x,\lambda(t))$ at the beginning $t=0$ and end $t=\tau$ of the driving process, we impose boundary conditions
\begin{equation}\dot{\lambda}(0) = \dot{\lambda}(\tau) = 0.\label{Eq-overdamped-boundary}\end{equation}
Combining the auxiliary potential (\ref{Eq-overdamped-U1solution}) and the boundary conditions (\ref{Eq-overdamped-boundary}), we can achieve shortcuts to isothermality in the overdamped situation.

As an illustration, we consider two simple, but frequently used examples. First, consider a Brownian particle moving in a harmonic potential and dragged by a time-dependent force. The corresponding potential reads
\begin{equation}U_{0}(x,\lambda(t))=\frac{1}{2}k x^{2} - \lambda(t) x\label{Eq-overdamped-harmonicpotential1}\end{equation}
where $k$ represents the constant stiffness of the harmonic potential and $\lambda(t)$ the external dragging force. Substituting Eq.~(\ref{Eq-overdamped-harmonicpotential1}) into Eq.~(\ref{Eq-overdamped-U1solution}), we obtain the auxiliary potential
\begin{equation}U_{1}(x,t)=-\frac{\gamma\dot{\lambda}(t)}{k} x.\label{Eq-overdamped-harmonicU1potential1}\end{equation}
Second, consider the motion of a Brownian particle in a time-dependent harmonic potential:
\begin{equation}U_{0}(x,\lambda(t))=\frac{1}{2}\lambda(t) x^{2}.\label{Eq-overdamped-harmonicpotential2}\end{equation}
Substituting Eq.~(\ref{Eq-overdamped-harmonicpotential2}) into Eq.~(\ref{Eq-overdamped-U1solution}), we obtain the auxiliary potential
\begin{equation}U_{1}(x,t)=\frac{\gamma\dot{\lambda}(t)}{4\lambda(t)} x^{2}.\label{Eq-overdamped-harmonicU1potential2}\end{equation}
The result equivalent to Eq.~(\ref{Eq-overdamped-harmonicU1potential2}) has been obtained in Ref.~\cite{Martinez2015} (see Eq.~(6) therein). In addition, their results imply that one can find analytical solutions of the auxiliary potential for a class of potentials, i.e., $U_{0}(x,\lambda(t)) = \lambda(t) x^{n}/2~(n=2,4,6 \cdots)$. The corresponding auxiliary potentials are $U_{1}(x,t) = \gamma \dot{\lambda}x^{2}/(2n \lambda)$ (see discussion below Eq.~(2) of the supplementary information in Ref.~\cite{Martinez2015}). Besides, it is easy to find auxiliary potentials for another class of potentials $U_{0}(x,\lambda(t)) = u(x -\lambda(t))$ which represent moving potentials with the same profile of the function $u=u(x)$. The corresponding auxiliary potentials can be explicitly expressed as $U_{1}(x,t) = - \gamma \dot{\lambda} x$. The detailed derivations of the above two broad classes of auxiliary potentials are shown in Appendix \ref{Sec-one}.

\section{Shortcuts to isothermality for a Brownian particle in the underdamped situation}
In the underdamped situation, the inertial effect of the Brownian particle plays an important role. We assume that the auxiliary potential $U_{1}$ is a function of the coordinate $x$ and the momentum $p$ of the particle~\cite{auxiliaryH}. Then the total Hamiltonian in this situation can be expressed as
\begin{equation}H(x,p,t) = H_{0}(x,p,\lambda(t)) + U_{1}(x,p,t)  ,\label{Eq-underdamped-Hamiltonian}\end{equation}
where $H_{0}(x,p,\lambda(t))= \frac{p^{2}}{2} + U_{0}(x,\lambda(t))$. For the sake of simplicity, we have set the mass of the particle being unit. The famous Kramers equation describes the time evolution of the distribution function $\rho(x,p,t)$ of the particle while the external potential is independent of the momentum $p$. By altering the Langevin equation that governs the motion of the particle, we generalize the Kramers equation to the case where the potential contains momentum $p$,
\begin{eqnarray}\frac{\partial \rho}{\partial t} = && -\frac{\partial}{\partial x} (p\rho) + \frac{\partial}{\partial p}\left( \gamma p\rho + \rho \frac{\partial U_{0}}{\partial x} \right)+ \frac{\gamma}{ \beta} \frac{\partial^{2} \rho}{\partial p^{2}} \nonumber \\ && - \frac{\partial U_{1} }{\partial p} \frac{\partial \rho}{\partial x} + \frac{\partial U_{1} }{\partial x} \frac{\partial \rho}{\partial p} + \gamma\frac{\partial}{\partial p}\left( \rho \frac{\partial U_{1} }{\partial p}\right). \label{Eq-underdamped-FPequation}\end{eqnarray}
The detailed derivation of the above equation is attached in Appendix \ref{Sec-two}.
The instantaneous equilibrium distribution of the original system follows
\begin{equation}\rho(x,p,t)=\rho_{\mathrm{ieq}}(x,p,\lambda(t)) \equiv \mathrm{e}^{\beta [F(\lambda(t))-H_{0}(x,p,\lambda(t))]}\label{Eq-underdamped-equilibrium}\end{equation}
with $F(\lambda) \equiv -\beta^{-1}\ln [\int_{-\infty}^{+\infty}\mathrm{d}x  \int_{-\infty}^{+\infty} \mathrm{d}p \mathrm{e}^{-\beta H_{0}(x,p,\lambda)}]$. Substituting Eq.~(\ref{Eq-underdamped-equilibrium}) into Eq.~(\ref{Eq-underdamped-FPequation}), we obtain
\begin{equation} \frac{\gamma}{\beta } \frac{\partial^{2}U_{1}}{\partial p^{2}} - \gamma p \frac{\partial U_{1}}{\partial p} +\frac{\partial U_{0}}{\partial x}\frac{\partial U_{1}}{\partial p} - p \frac{\partial U_{1}}{\partial x}   =\left ( \frac{\mathrm{d} F}{\mathrm{d} \lambda} - \frac{\partial U_{0}}{\partial \lambda} \right ) \dot{\lambda}   . \label{Eq-underdamped-H1limitation}\end{equation}
By choosing
\begin{equation}U_{1}(x,p,t)= \dot{\lambda}(t)f(x,p,\lambda(t)),\label{Eq-underdamped-H1form}\end{equation}
we can derive a partial differential equation
\begin{equation}  \frac{\gamma}{\beta} \frac{\partial^{2}f}{\partial p^{2}} - \gamma p \frac{\partial f}{\partial p}  + \frac{\partial U_{0}}{\partial x}\frac{\partial f}{\partial p} - p \frac{\partial f}{\partial x}   = \frac{\mathrm{d} F}{\mathrm{d} \lambda} - \frac{\partial U_{0}}{\partial \lambda}   . \label{Eq-underdamped-flimitation}\end{equation}
It seems unlikely to solve this equation for $f(x,p,\lambda(t))$ analytically, except for a few simple systems.

Here, we still consider the two special potentials (\ref{Eq-overdamped-harmonicpotential1}) and (\ref{Eq-overdamped-harmonicpotential2}). Correspondingly, we obtain auxiliary potentials
\begin{equation} U_{1}(x,p,t) =\frac{\dot{\lambda}(t)}{k} (p - \gamma x)\label{Eq-underdamped-harmonicH1form1}\end{equation}
and
\begin{equation} U_{1}(x,p,t) =\frac{\dot{\lambda}(t)}{4\gamma\lambda(t)} [ (p - \gamma x)^{2}+ \lambda(t)x^{2}].\label{Eq-underdamped-harmonicH1form2}\end{equation}
Analogous to the overdamped situation, we also find analytical solutions of the auxiliary potential for the two classes of potentials, $U_{0}(x,\lambda(t)) = \lambda(t) x^{n}/2~(n=2,4,6 \cdots)$ and $U_{0}(x,\lambda(t)) = u(x -\lambda(t))$, in the underdamped situation. The corresponding auxiliary potentials read
$U_{1}(x,p,t) =\dot{\lambda} [ (p - \gamma x)^{2}+ \lambda(t)x^{n}]/2n\gamma\lambda$ and $ U_{1}(x,p,t) =\dot{\lambda} (p - \gamma x)$, respectively. The detailed derivations of the above two broad classes of auxiliary potentials are shown in Appendix \ref{Sec-one}.

%As a consistency check, we find that in the overdamped limit, auxiliary potentials (\ref{Eq-underdamped-harmonicH1form1}) and (\ref{Eq-underdamped-harmonicH1form2}) reduce to (\ref{Eq-overdamped-harmonicU1potential1}) and (\ref{Eq-overdamped-harmonicU1potential2}), respectively.
The experimental work by Le Cunuder and his coworkers~\cite{Cunuder2016} suggests that the auxiliary potential~(\ref{Eq-underdamped-harmonicH1form1}) can be realized in laboratory.
The cross term $xp$ in above auxiliary potential~(\ref{Eq-underdamped-harmonicH1form2}) resembles the auxiliary counterdiabatic Hamiltonian in shortcuts to adiabaticity~\cite{Jarzynski2013,Campo2013,Deffner2014}. A possible experimental scheme was proposed by del Campo to realize such a cross term for a broad family of many-body quantum systems controlled by shortcuts to adiabaticity~\cite{Campo2013}. The requirement for detailed instantaneous microscopic knowledge is completely removed for such systems. With proper canonical transformation, he derived an alternative representation of the counterdiabatic Hamiltonian in which the cross term is absent. His scheme may provide a clue to realize auxiliary potentials~(\ref{Eq-underdamped-harmonicH1form2}) for the system of Brownian particles.

\section{Nonequilibrium work relations}
In this section, we will theoretically derive three nonequilibrium work relations when adapting shortcuts to isothermality for Brownian particles moving in general potentials.

\subsection{Relation between the free energy difference and the intrinsic work}
Now we discuss the relation between the free energy difference and work.

By taking a derivative of the free energy (\ref{Eq-freeenergy}) with respect to $\lambda$,
one can obtain
\begin{equation}    \frac{\mathrm{d} F(\lambda)}{\mathrm{d} \lambda} = \left \langle \frac{\partial H_{0}}{\partial \lambda} \right \rangle. \label{Eq-differentiate-freeenergy}\end{equation}
Hereinafter, $\langle \cdots \rangle$ represents the average with the instantaneous equilibrium distribution function (\ref{Eq-instant-equilibriumDF}) since the real distribution of the system controlled by shortcuts to isothermality is exactly equal to (\ref{Eq-instant-equilibriumDF}). Multiplying both sides of Eq.~(\ref{Eq-differentiate-freeenergy}) by $\dot{\lambda}$ and integrating over time, we obtain
\begin{equation} \Delta F =  \int^{\tau}_{0}\mathrm{d}t \dot{\lambda} \left \langle \frac{\partial H_{0}}{\partial \lambda} \right \rangle. \label{Eq-freeenergygeneral}\end{equation}
For a Brownian particle moving in a time-dependent potential $U_{0}(x,\lambda(t))$, Eq.~(\ref{Eq-freeenergygeneral}) reduces to
\begin{equation} \Delta F =  \int^{\tau}_{0}\mathrm{d}t \dot{\lambda} \left \langle \frac{\partial U_{0}}{\partial \lambda} \right \rangle. \label{Eq-freeenergy2}\end{equation}
Equation (\ref{Eq-freeenergy2}) indicates that the free energy difference is merely determined by the mean work related to the original Hamiltonian of the system (we dub it intrinsic work). This equation holds for finite-rate nonequilibrium processes with the adaption of shortcuts to isothermality. Without the adaption of the auxiliary potential, Eq.~(\ref{Eq-freeenergy2}) holds only for quasi-static processes.

\subsection{Relation between the dissipated work and the driving time}
In this subsection, we investigate the relation between the dissipated work and the driving time.

According to stochastic thermodynamics~\cite{Sekimoto2010,Seifert2012}, the mean work done in the shortcut to isothermality driving process follows
\begin{eqnarray}W && = \int^{\tau}_{0}\mathrm{d}t \left\langle  \frac{\partial H}{\partial t} \right \rangle \nonumber \\ && = \int^{\tau}_{0}\mathrm{d}t \left( \dot{\lambda}  \left\langle \frac{\partial H_{0}}{\partial \lambda} \right\rangle + \dot{\lambda}  \left\langle \frac{\partial U_{1}}{\partial \lambda} \right\rangle + \ddot{\lambda}  \left\langle \frac{\partial U_{1}}{\partial \dot{\lambda}} \right\rangle \right).~~ \label{Eq-overdamped-meanwork}\end{eqnarray}
From Eqs.~(\ref{Eq-freeenergygeneral}) and (\ref{Eq-overdamped-meanwork}), the dissipated work in this driving process may be expressed as
\begin{equation} W_{\mathrm{d}} \equiv W - \Delta F  =   \int^{\tau}_{0}\mathrm{d}t \left( \dot{\lambda}  \left\langle \frac{\partial U_{1}}{\partial \lambda} \right\rangle + \ddot{\lambda}  \left\langle \frac{\partial U_{1}}{\partial \dot{\lambda}} \right\rangle \right).  \label{Eq-general-meanirrwork}\end{equation}
The non-negativity of the dissipated work is shown in Appendix \ref{Sec-three}.
Substituting Eq.~(\ref{Eq-overdamped-U1form}) or Eq.~(\ref{Eq-underdamped-H1form}) into Eq.~(\ref{Eq-general-meanirrwork}), we obtain
\begin{equation} W_{\mathrm{d}} = \int^{\tau}_{0}\mathrm{d}t \left( \dot{\lambda}^{2} \left\langle  \frac{\partial f}{\partial \lambda} \right\rangle  +  \ddot{\lambda} \left\langle  f \right\rangle  \right).  \label{Eq-general-meanwork}\end{equation}
Since the controlling parameter is fixed at the beginning and end of the driving process, it will be natural to choose the protocol with the reducible form, $\lambda(t) = \lambda(t/ \tau)$. Through rescaling the time $s\equiv t/ \tau$, we can reformulate Eq.~(\ref{Eq-general-meanwork}) as
\begin{equation}  W_{\mathrm{d}}  =  \frac{1}{\tau}   \int^{1}_{0}\mathrm{d}s \left[ \left( \frac{\mathrm{d} \lambda}{\mathrm{d}s} \right)^{2} \left\langle  \frac{\partial f}{\partial \lambda} \right\rangle  +  \frac{\mathrm{d}^{2} \lambda}{\mathrm{d} s^{2}} \left\langle  f \right\rangle  \right] .              \label{Eq-general-meanwork2}\end{equation}
The numerator of Eq.~(\ref{Eq-general-meanwork2}) in this case is independent of the driving time $\tau$, which reveals that the dissipated work $W_{\mathrm{d}}$ (with shortcuts to isothermality being adapted) is inversely proportional to the driving time $\tau$. This coincides with the conclusion obtained by Schmiedl and Seifert~\cite{Schmiedl2008} who found that, in the overdamped situation, the dissipated work done during the optimal driving process always scales with the inverse driving time. Moreover, we surprisingly find that the dissipated work (with the adaption of shortcuts to isothermality) is still proportional to the inverse driving time in the overdamped and underdamped situations under a reducible driving protocol.

\subsection{Jarzynski-like equality with shortcuts to isothermality}
In this subsection, we will derive a Jarzynski-like equality when adapting shortcuts to isothermality.

Crooks~\cite{Crooks2000} proposed a theorem
\begin{equation}   \left\langle  \mathcal{O}   \right\rangle_{+} =   \langle  \bar{\mathcal{O}} \mathrm{e}^{-\beta(w-\Delta F)}  \rangle_{-},\label{Eq-Crooks-theorem}\end{equation}
which holds true for a system departing from an equilibrium state driven by an external field in finite rate.
Here, $\mathcal{O}$ is a functional of the trajectory, while $\bar{\mathcal{O}}$ is the corresponding time-reversed counterpart. $w$ is the work performed on the system along a single trajectory. $\langle \cdots \rangle_{+}$ represents the ensemble average over trajectories stemming from initial equilibrium state in the forward driving process. $\langle \cdots \rangle_{-}$ represents the ensemble average over trajectories stemming from an equilibrium state in the time-reversal driving process.
In particular, if one takes $\mathcal{O}=\delta[\Gamma-\Gamma(\tau)]$, the above theorem (\ref{Eq-Crooks-theorem}) will lead to~\cite{Sivak2012,Gong2015}:
\begin{eqnarray}  \rho_{+}(\Gamma,\tau)  && =  \langle  \delta[\Gamma-\Gamma(\tau)] \rangle_{+} \nonumber \\  && = \langle  \delta[\Gamma-\bar{\Gamma}(\tau)] \mathrm{e}^{-\beta(w-\Delta F)} \rangle_{-} \nonumber \\  && =  \rho_{\mathrm{eq}}(\Gamma) \langle \mathrm{e}^{-\beta(w-\Delta F)} \rangle_{\Gamma,-} .\label{Eq-Kawasaki-relation}\end{eqnarray}
The subscript $``\Gamma,-"$ indicates the ensemble average over all trajectories starting from a fixed state $\Gamma$ in the time-reversal driving process. Equation (\ref{Eq-Kawasaki-relation}) describes the relationship between the distribution function $\rho_{+}(\Gamma,\tau)$ of final states in the forward driving process and the corresponding equilibrium distribution $\rho_{\mathrm{eq}}(\Gamma)$ when the protocol is fixed at $\lambda_\tau$. By implementing the strategy of shortcuts to isothermality in the forward driving process, we can evolve the system from an equilibrium state to another one at the same temperature, which means $\rho_{+}(\Gamma,\tau)=\rho_{\mathrm{eq}}(\Gamma)$ in Eq.~(\ref{Eq-Kawasaki-relation}). Thus we obtain
\begin{equation}  \langle \mathrm{e}^{-\beta (w-\Delta F)} \rangle_{\Gamma,-} = 1.  \label{Eq-third-property}\end{equation}
The above equation follows from Eq.~(\ref{Eq-Crooks-theorem}) which is based on an assumption that the system Hamiltonian $H(\Gamma,t)$ is time-reversal invariant~\cite{Crooks2000}. This assumption is obviously valid for shortcuts to isothermality in the overdamped situation. Nonetheless, one should pay attention to the subtle point in the underdamped situation, where in the reversed protocol, one should reverse the sign of all the momenta.

Equation (\ref{Eq-third-property}) implies that we can estimate $\Delta F$ by taking the exponential average of $w$, over trajectories that start from a fixed state $\Gamma$ and then evolve under the time-reversal of the forward protocol with the adaptation of shortcuts to isothermality. If we further multiply an arbitrary distribution $\rho (\Gamma)$ to both sides of Eq.~(\ref{Eq-third-property}) and perform the integral over the whole phase space, we may obtain $\int\rho (\Gamma) \langle \mathrm{e}^{-\beta (w-\Delta F)} \rangle_{\Gamma,-} \mathrm{d} \Gamma = 1$, which can also be obtained from a result in Ref.~\cite{Gong2015}.

Equation (\ref{Eq-third-property}) has the similar form as the Jarzynski equality
\begin{equation}  \langle \mathrm{e}^{-\beta (w-\Delta F)} \rangle = 1,  \label{Eq-traditional-JE}\end{equation}
thus we dub Eq.~(\ref{Eq-third-property}) Jarzynski-like equality.
The ensemble average in the Jarzynski equality is made over trajectories starting from an equilibrium distribution while the ensemble average in the Jarzynski-like equality is made over trajectories starting from a fixed state $\Gamma$ and then evolve under the time-reversal of the forward protocol with the adaptation of shortcuts to isothermality. Besides, the Jarzynski-like equality requires an additional constraint for the driving protocol, namely that the driving protocol should be designed according to the strategy of shortcuts to isothermality.

\section{Numerical confirmations}
To illustrate above relations (\ref{Eq-freeenergy2}) and (\ref{Eq-general-meanwork2}), we simulate the overdamped motion of a Brownian particle in the potential (\ref{Eq-overdamped-harmonicpotential1}) and add the corresponding auxiliary potential (\ref{Eq-overdamped-harmonicU1potential1}) to realize shortcuts to isothermality. Since relation (\ref{Eq-third-property}) involves time-reversed trajectories, we need to use the minus auxiliary potential to compute the trajectory work (see Appendix \ref{Sec-four} for the detailed derivation).

The overdamped motion of a Brownian particle in the potential
\begin{eqnarray}    U(x,t) &=& U_{0}(x,\lambda(t)) + U_{1}(x,t) \nonumber \\ &=& \frac{1}{2}k x^{2} - \lambda(t) x -\frac{\gamma\dot{\lambda}(t)}{k} x \label{Eq-total-potential}\end{eqnarray}
is governed by the Langevin equation
\begin{equation}    \gamma \dot{x} = - k x + \lambda + \frac{\gamma\dot{\lambda}}{k} +  \xi(t). \label{Eq-overdamped-LE}\end{equation}
Introduce the characteristic length $l_{c} \equiv \sqrt{k_{B}T/k}$ and the characteristic time $\tau_{c} \equiv \gamma / k$ of the system, and then reduce the coordinate $\tilde{x} \equiv x / l_{c}$ , the time $s \equiv t / \tau$ and the driving protocol $\tilde{\lambda} \equiv \lambda / (kl_{c})$. The above Langevin equation may be transformed into the following dimensionless form:
\begin{equation}    \tilde{x}'(s) = - \tilde{\tau} \tilde{x}(s) + \tilde{\tau} \tilde{\lambda}(s) + \tilde{\lambda}'(s) + \sqrt{2\tilde{\tau}} \zeta(s) \label{Eq-dimensionless-LE}\end{equation}
with $\tilde{\tau} \equiv \tau / \tau_{c} = k\tau/\gamma$. The prime on a variable represents the derivative of that variable with respect to the normalized time $s$. $\zeta(s)$ represents Gaussian white noise that satisfies $\langle \zeta(s) \rangle = 0$ and $\langle \zeta(s_{1}) \zeta(s_{2}) \rangle = \delta(s_{1}-s_{2})$. Equation~(\ref{Eq-dimensionless-LE}) was solved using the Euler algorithm
\begin{equation}    \tilde{x}(s+\delta s) = \tilde{x}(s) - \tilde{\tau} \tilde{x}(s)\delta s + \tilde{\tau} \tilde{\lambda}(s)\delta s + \tilde{\lambda}'(s)\delta s + \sqrt{2\tilde{\tau}\delta s} \theta (s) \label{Eq-Euler-algorithm}\end{equation}
where $\delta s $ is the time step. $\theta(s)$ is a random number sampled from Gaussian distribution with zero mean and unit variance.

We choose the protocol evolving in the form
\begin{equation}    \tilde{\lambda}(s) = 4[ 1-  \mathrm{cos}(\pi s)],~~~~0\le s \le 1, \label{Eq-protocol}\end{equation}
and set $k_{B}T=1$ and the time step $\delta s = 10^{-5}$. The theoretical value of the free energy difference corresponding to the above driving protocol is $\Delta F = -32$ ($k_{B}T$). We obtain each value of $\Delta F$ by using $10^{6}$ trajectories, evolving under the dimensionless Langevin equation (\ref{Eq-dimensionless-LE}).

\begin{figure}[!htp]
%\centering
 \includegraphics[width = 7cm]{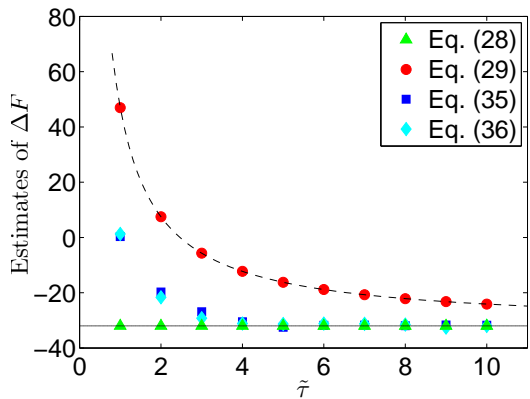}
 %\captionsetup{font={scriptsize}}
 \caption{\label{fig1}(Color online). Comparison of estimates of $\Delta F$ from Eqs.~(\ref{Eq-freeenergy2}) (triangles), (\ref{Eq-overdamped-meanwork}) (circles), (\ref{Eq-third-property}) (squares), and the Jarzynski equality (\ref{Eq-traditional-JE}) (diamonds). The dashed line is a fitted curve for Eq.~(\ref{Eq-overdamped-meanwork}) with the consideration of Eq.~(\ref{Eq-general-meanwork2}), while the solid line represents the theoretical value for $\Delta F =-32$.}
\end{figure}

We perform simulations for different driving times ranging from $\tilde{\tau}=1.0$ to $\tilde{\tau}=10.0$. The initial state of trajectories is sampled from the equilibrium distribution when we estimate the free energy difference by using Eqs.~(\ref{Eq-freeenergy2}), (\ref{Eq-overdamped-meanwork}), and the Jarzynski equality (\ref{Eq-traditional-JE}). As for Eq.~(\ref{Eq-third-property}), the initial state is fixed at $\Gamma=0.0$ and the time-reversed trajectories are generated by the Langevin dynamics $\gamma \dot{x} = - k x + \lambda - \gamma\dot{\lambda}/k +  \xi(t)$. Figure \ref{fig1} shows the results of estimates of $\Delta F$ from Eqs.~(\ref{Eq-freeenergy2}), (\ref{Eq-overdamped-meanwork}), (\ref{Eq-third-property}), and (\ref{Eq-traditional-JE}), respectively. The estimates of $\Delta F$ from Eqs.~(\ref{Eq-third-property}) and (\ref{Eq-traditional-JE}) converge to the theoretical value much faster than the mean work (\ref{Eq-overdamped-meanwork}). The estimated results from Eq.~(\ref{Eq-freeenergy2}) are remarkably accurate over the entire range of driving times.

\begin{figure}[!htb]
%\centering
 \includegraphics[width = 7cm]{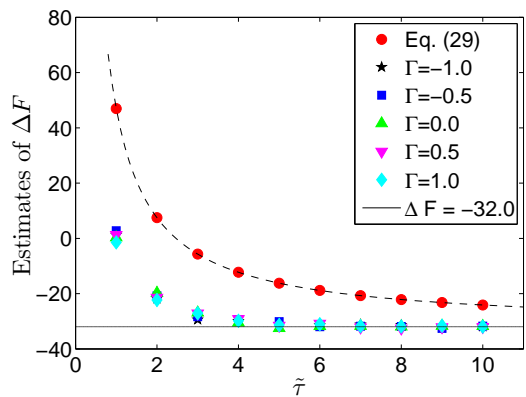}
 %\captionsetup{font={scriptsize}}
 \caption{\label{fig2}(Color online). Comparison of estimates of $\Delta F$ from relation (\ref{Eq-third-property}) with different initial fixed states.}

\end{figure}

We also compare estimates of $\Delta F$ from relation (\ref{Eq-third-property}) for trajectories starting from different initial states $\Gamma \equiv \tilde{x}(0)=  -1.0$, -0.5, 0.0, 0.5, 1.0, respectively. The values of fixed states are selected according to the standard deviation of the initial equilibrium distribution, $\rho(\tilde{x},0) \propto \mathrm{e}^{-\tilde{x}^{2}/2} $. Figure \ref{fig2} shows that the estimates of $\Delta F$ for different initial fixed states are very similar to each other in relation (\ref{Eq-third-property}). This confirms that the free energy difference can be extracted from an ensemble of trajectories starting from a fixed state with the adaptation of shortcuts to isothermality.

\begin{figure}[!htb]
%\centering
 \includegraphics[width = 7cm]{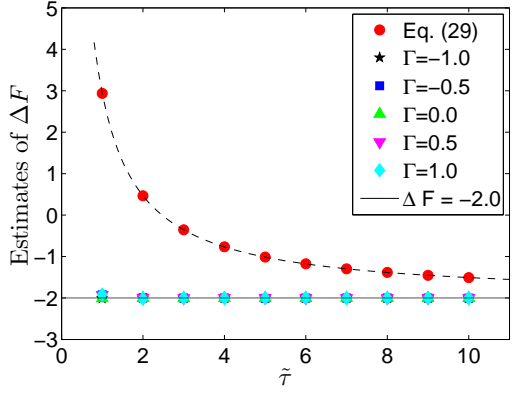}
 %\captionsetup{font={scriptsize}}
 \caption{\label{fig3}(Color online). Comparison of estimates of $\Delta F$ from relation (\ref{Eq-third-property}) with different initial fixed states. The process is conducted under a new driving protocol, $\tilde{\lambda}(s) =  1-  \mathrm{cos}(\pi s)$. }

\end{figure}

Note that for short driving times, relations (\ref{Eq-third-property}) and (\ref{Eq-traditional-JE}) lead to relative large errors when we use them to estimate the free energy difference $\Delta F$. This deviation is due to the significant magnitude of dissipated work for the short driving process. Jarzynski found that the number of trajectories required for convergence of the traditional Jarzynski equality grows exponentially in the dissipated work~\cite{Jarzynski2006}. If we suppress the dissipated work by replacing protocol (\ref{Eq-protocol}) with $\tilde{\lambda}(s) =  1-  \mathrm{cos}(\pi s)$, relative errors for fast driving process will be reduced. As shown in Figure \ref{fig3}, all values of the free energy difference estimated from relation (\ref{Eq-third-property}) for trajectories starting from different initial states $\Gamma \equiv \tilde{x}(0)=  -1.0$, -0.5, 0.0, 0.5 and 1.0 approach the theoretical result $\Delta F=-2$.

\section{Conclusion}
In the discussion above, we proposed the concept of shortcuts to isothermality and constructed the strategy to explicitly realize the transition between two equilibrium states at the same temperature in finite time. Our work provides a unified theoretical framework for designing engineered swift equilibration~\cite{Martinez2015,Cunuder2016}.

We have also found three nonequilibrium work relations when adapting shortcuts to isothermality. Equation (\ref{Eq-freeenergy2}) implies that we can estimate the free energy difference by calculating the average work due to the time-dependent potential of the original system, which may offer an efficient method to evaluate the free energy difference in simulations. Equation (\ref{Eq-general-meanwork2}) implies that the dissipated work when adapting shortcuts to isothermality is inversely proportional to the driving time under a reducible driving protocol, which generalizes the result put forward by Schmiedl and Seifert~\cite{Schmiedl2008} in their overdamped model of stochastic heat engines. Equation (\ref{Eq-third-property}) is a Jarzynski-like equality for open classical systems, which relaxes the requirement for initial distributions. These relations may be applied in estimates of the free energy in simulations of protein folding and experiments of single molecule mechanics.

Finally, there seems no obstacle to implement shortcuts to isothermality experimentally in the overdamped situation since the auxiliary potential (\ref{Eq-overdamped-U1solution}) depends merely on the protocol $\lambda(t)$ and coordinate $x$. In particular, auxiliary potentials (\ref{Eq-overdamped-harmonicU1potential1}) and (\ref{Eq-overdamped-harmonicU1potential2}) are easily achieved in experiments. The big challenge is the realization of shortcuts to isothermality in the underdamped situation, which summons future collaborations between theoretical and experimental researchers. Besides, we could construct finite-time thermodynamic cycles with a combination of shortcuts to adiabaticity and shortcuts to isothermality. In these cycles, the distribution function of the system is well-determined at any time, which might overcome the shortage of traditional models of finite-rate heat engines.

\emph{Acknowledgement.}--The authors are grateful to financial support from the
National Natural Science Foundation of China (Grant NOs. 11675017, 11322543, 11375012, and 11534002) and the Fundamental Research Funds for the Central Universities (NOs. 2015KJJCB01, 2017EYT24, 2017STUD23). They also acknowledge valuable discussions with J. Bechhoefer, E. Trizac, J. G. Bao, D. J. Searles, H. Qian, P. Ao, M. Esposito, X. Zhou, and X. R. Ma.

\appendix
\section{Detailed derivations of auxiliary potentials\label{Sec-one}}

In this section, we will derive auxiliary potentials for two broad classes of potentials, $U_{0}(x,\lambda(t)) = \lambda(t) x^{n}/2~(n=2,4,6 \cdots)$ and $U_{0}(x,\lambda(t)) = u(x -\lambda(t))$, in the overdamped and underdamped situations respectively.

\subsection{Overdamped Situation}

In the overdamped situation, the instantaneous equilibrium distribution corresponding to potentials $U_{0}(x,\lambda(t)) = \lambda(t) x^{n}/2~(n=2,4,6 \cdots)$ reads \begin{equation}\rho(x,t)= \rho_{\mathrm{ieq}}(x,\lambda(t)) \equiv \mathrm{e}^{\beta \left [ F(\lambda(t))- \frac{1}{2}\lambda(t) x^{n} \right]},\label{Eq-instant-equilibriumDFO1}\end{equation}
with
\begin{eqnarray} F(\lambda) \equiv -\beta^{-1}\ln \int_{-\infty}^{+\infty} \mathrm{e}^{-\frac{1}{2}\beta \lambda x^{n} }\mathrm{d}x   = -\beta^{-1}\ln \left( C \lambda^{-\frac{1}{n}} \right) . \label{Eq-freeenergyO1}\end{eqnarray}
Here, $C=\int_{-\infty}^{+\infty} \mathrm{e}^{-\frac{1}{2}\beta x^{n} }\mathrm{d}x$ is constant. Substituting Eqs.~(\ref{Eq-instant-equilibriumDFO1}) and ~(\ref{Eq-freeenergyO1}) into Eq.~(\ref{Eq-overdamped-U1solution}), we obtain auxiliary potentials:
\begin{eqnarray}  U_{1}(x,t)&&= \gamma \beta \dot{\lambda} \int_{0}^{x}\mathrm{d}y\frac{\int_{-\infty}^{y}\mathrm{d} z \left[\left( \frac{1}{n\beta \lambda} - \frac{1}{2}z^{n} \right) \mathrm{e}^{-\frac{1}{2}\beta \lambda z^{n} } \right]}{\mathrm{e}^{-\frac{1}{2}\beta \lambda y^{n} }}  \nonumber \\ &&  =\gamma \beta \dot{\lambda}  \int_{0}^{x}\mathrm{d}y\frac{\frac{1}{ n \beta \lambda} \int_{-\infty}^{y}\mathrm{d} z     \left [  \frac{\mathrm{d}}{\mathrm{d} z}\left( z \mathrm{e}^{-\frac{1}{2}\beta \lambda z^{n} } \right) \right]}{\mathrm{e}^{-\frac{1}{2}\beta \lambda y^{n} }}  \nonumber \\ &&   = \gamma \beta \dot{\lambda} \int_{0}^{x}\mathrm{d}y \frac{y}{n \beta \lambda} = \frac{\gamma \dot{\lambda}}{2n \lambda} x^{2}. \label{Eq-auxiliaryPO1}\end{eqnarray}
Hereinafter, in order to obtain a concise expression of auxiliary potentials, we choose specific lower bounds for integrals.

The instantaneous equilibrium distribution corresponding to potentials $U_{0}(x,\lambda(t)) = u(x -\lambda(t))$ reads
\begin{equation}\rho(x,t)= \rho_{\mathrm{ieq}}(x,\lambda(t)) \equiv \mathrm{e}^{\beta \left [ F- u(x -\lambda(t)) \right]},\label{Eq-instant-equilibriumDFO2}\end{equation}
where $F  \equiv -\beta^{-1}\ln \left[\int_{-\infty}^{+\infty} \mathrm{e}^{-\beta u(x -\lambda) }\mathrm{d}x \right]$ is constant. Substituting Eqs.~(\ref{Eq-instant-equilibriumDFO2}) into Eq.~(\ref{Eq-overdamped-U1solution}), we obtain auxiliary potentials:
\begin{eqnarray}  U_{1}(x,t)&&= \gamma \beta \dot{\lambda} \int_{0}^{x}\mathrm{d}y\frac{\int_{-\infty}^{y}\mathrm{d} z \left[ - \frac{\partial u}{\partial \lambda}  \mathrm{e}^{-\beta u(z -\lambda(t)) } \right]}{\mathrm{e}^{-\beta u(y-\lambda(t)) }}  \nonumber \\ && = \gamma \beta \dot{\lambda} \int_{0}^{x}\mathrm{d}y\frac{\int_{-\infty}^{y}\mathrm{d} (z -\lambda(t)) \left[\frac{\mathrm{d} u}{\mathrm{d} (z -\lambda(t))}  \mathrm{e}^{-\beta u(z -\lambda(t))} \right]}{\mathrm{e}^{-\beta u(y -\lambda(t)) }}  \nonumber \\ && = \gamma \beta \dot{\lambda} \int_{0}^{x}\mathrm{d}y\frac{\int_{-\infty}^{y}\mathrm{d} q \left[\frac{\mathrm{d} u}{\mathrm{d} q}  \mathrm{e}^{-\beta u(q )} \right]}{\mathrm{e}^{-\beta u(y -\lambda(t)) }}  = -\gamma \dot{\lambda} x. \label{Eq-auxiliaryPO2}\end{eqnarray}

\subsection{Underdamped Situation}

In the underdamped situation, the instantaneous equilibrium distribution corresponding to potentials $U_{0}(x,\lambda(t)) = \lambda(t) x^{n}/2~(n=2,4,6 \cdots)$ reads \begin{equation}\rho(x,p,t)= \rho_{\mathrm{ieq}}(x,p,\lambda(t)) \equiv \mathrm{e}^{\beta \left [ F(\lambda(t))- \frac{p^{2}}{2}-\frac{1}{2}\lambda(t) x^{n} \right]},\label{Eq-instant-equilibriumDFU1}\end{equation}
with
\begin{eqnarray} F(\lambda) && \equiv -\beta^{-1}\ln \left(  \int_{-\infty}^{+\infty} \mathrm{e}^{-\frac{1}{2}\beta p^{2} }\mathrm{d}p   \int_{-\infty}^{+\infty} \mathrm{e}^{-\frac{1}{2}\beta \lambda x^{n} }\mathrm{d}x \right) \nonumber \\ && = -\beta^{-1}\ln \left( C \lambda^{-\frac{1}{n}} \right) . \label{Eq-freeenergyU1}\end{eqnarray}
Here, $C=\int_{-\infty}^{+\infty} \mathrm{e}^{-\frac{1}{2}\beta p^{2} }\mathrm{d}p \int_{-\infty}^{+\infty} \mathrm{e}^{-\frac{1}{2}\beta x^{n} }\mathrm{d}x$ is constant.

We assume the function $f(x,p,\lambda(t))$ to be a polynomial, $f(x,p, \lambda) = a_{1}(\lambda)p^{2} + a_{2}(\lambda)xp + a_{3}(\lambda)x^{2}+a_{4}(\lambda)x^{n}$. Substituting it and Eq.~(\ref{Eq-freeenergyU1}) into Eq.~(\ref{Eq-underdamped-flimitation}), we obtain
\begin{equation} f(x,p,t) =\frac{1}{2n\gamma \lambda}p^{2} -\frac{1}{n\lambda}xp + \frac{\gamma}{2n\lambda}x^{2} + \frac{1}{2n\gamma}x^{n}.     \label{Eq-auxiliaryPU1}\end{equation}
Combining Eq.~(\ref{Eq-auxiliaryPU1}) with Eq.~(\ref{Eq-underdamped-H1form}), we can derive the corresponding auxiliary potentials in the main text.

The instantaneous equilibrium distribution corresponding to potentials $U_{0}(x,\lambda(t)) = u(x -\lambda(t))$ reads
\begin{equation}\rho(x,p,t)= \rho_{\mathrm{ieq}}(x,p,\lambda(t)) \equiv \mathrm{e}^{\beta \left [ F- \frac{p^{2}}{2}-u(x -\lambda(t)) \right]},\label{Eq-instant-equilibriumDFU2}\end{equation}
with
\begin{eqnarray} F \equiv -\beta^{-1}\ln \left[  \int_{-\infty}^{+\infty} \mathrm{e}^{-\frac{1}{2}\beta p^{2} }\mathrm{d}p   \int_{-\infty}^{+\infty} \mathrm{e}^{-\beta u(x -\lambda) }\mathrm{d}x \right] .  \label{Eq-freeenergyU2}\end{eqnarray}
It is not hard to prove that $F$ keeps constant.

We assume the function $f(x,p,\lambda(t))$ to be a linear polynomial, $f(x,p, \lambda) = a_{1}(\lambda)p + a_{2}(\lambda)x$. Substituting it and Eq.~(\ref{Eq-freeenergyU2}) into Eq.~(\ref{Eq-underdamped-flimitation}), we obtain
\begin{equation} f(x,p,t) = p - \gamma x .    \label{Eq-auxiliaryPU2}\end{equation}
Combining Eq.~(\ref{Eq-auxiliaryPU2}) with Eq.~(\ref{Eq-underdamped-H1form}), we can derive the corresponding auxiliary potentials in the main text.

\section{Extension of the Kramers equation\label{Sec-two}}

In the underdamped situation, the motion of a Brownian particle in a time-dependent potential $U_{0}(x,\lambda(t))$ is described by the Langevin equation~\cite{Langevin1908,Reichl1998}
\begin{equation} \dot{x}=p, \quad     \dot{p} = - \frac{\partial U_{0}(x,\lambda)}{\partial x} - \gamma p + \xi(t),\label{Eq-underdamped-LE}\end{equation}
where $\gamma$ is the coefficient of friction. For the sake of simplicity, we have set the mass of the particle being unit. $\xi(t)$ represents Gaussian white noise that satisfies $\langle \xi(t) \rangle = 0$ and $\langle \xi(t) \xi(t') \rangle = 2\gamma k_{B}T\delta(t-t')$. The time evolution of the distribution function $\rho(x,p,t)$ to observe the particle at position $x$ with momentum $p$ at time $t$ is then governed by the Kramers equation~\cite{Reichl1998}
\begin{equation} \frac{\partial \rho}{\partial t} = -\frac{\partial}{\partial x} (p\rho) + \frac{\partial}{\partial p}\left[\rho\left( \gamma p + \frac{\partial U_{0}}{\partial x} + \frac{\gamma}{\beta\rho}  \frac{\partial \rho}{\partial p} \right)\right].   \label{Eq-underdamped-FPE} \end{equation}
However, Eqs.~(\ref{Eq-underdamped-LE}) and (\ref{Eq-underdamped-FPE}) just deal with situations when the external potential is independent of the momentum $p$. If we introduce an auxiliary potential $U_{1}(x,p,t)$ that also depends on the momentum $p$ of the particle to the system, the above equations describing the evolution of the system need to be rebuilt. With the total Hamiltonian of the system (\ref{Eq-underdamped-Hamiltonian}), the corresponding canonical equations follow
\begin{equation} \begin{split}& \dot{x} =  \frac{\partial H}{\partial p}=p + \frac{\partial U_{1}}{\partial p}  \\ &     \dot{p}  =  - \frac{\partial H}{\partial x} = - \frac{\partial U_{0}}{\partial x} -  \frac{\partial U_{1}}{\partial x} . \end{split} \label{Eq-canonicalequation}\end{equation}
According to the pioneering work of Langevin~\cite{Langevin1908}, collisions with molecules around the Brownian particle can be treated as a viscous force $- \gamma \dot{x}$ and a random force $\xi(t)$. Therefore, the generalized Langevin equation that describing a Brownian particle in the momentum dependent potential $U_{0}(x,t) + U_{1}(x,p,t)$ becomes
\begin{equation} \begin{split}& \dot{x} = p + \frac{\partial U_{1}}{\partial p}   \\ &     \dot{p}  =  - \frac{\partial U_{0}}{\partial x} -  \frac{\partial U_{1}}{\partial x}  - \gamma p - \gamma \frac{\partial U_{1}}{\partial p} + \xi(t). \end{split}  \label{Eq-generalized-LE}\end{equation}
In general, Eq.~(\ref{Eq-generalized-LE}) can be treated as a two-variable stochastic differential equation~\cite{Gardiner1985}. The Fokker-Planck equation corresponding to it follows\begin{eqnarray}\label{Eq-generalized-FP}
         \frac{\partial \rho}{\partial t} &=& -\frac{\partial}{\partial x} \left [\rho \left(p + \frac{\partial U_{1}}{\partial p} \right) \right]   \\
          &+&  \frac{\partial}{\partial p}\left [\rho \left(\frac{\partial U_{0}}{\partial x}   + \frac{\partial U_{1}}{\partial x} + \gamma p + \gamma \frac{\partial U_{1}}{\partial p} +  \frac{\gamma}{ \beta \rho} \frac{\partial \rho}{\partial p}  \right) \right] \nonumber
       \end{eqnarray}
%\begin{equation} \begin{split} = & + \frac{\partial U_{1}}{\partial x} \right. \right.  \\ & \left. \left.+ \gamma p + \gamma \frac{\partial U_{1}}{\partial p} +  \frac{\gamma}{ \beta \rho} \frac{\partial \rho}{\partial p}  \right) \right] , \end{split}   \label{Eq-generalized-FP}\end{equation}
which is just the generalized Kramers equation (\ref{Eq-underdamped-FPequation}).

\section{Non-negativity of the Dissipated Work\label{Sec-three}}

In this section, we will discuss the non-negativity of the dissipated work respectively in the overdamped and underdamped situations.

\subsection{Overdamped Situation}

In the overdamped situation, Seifert~\cite{Seifert2005,Schmiedl2008} has derived the non-negative expression of the dissipated work for the Langevin dynamics,
\begin{equation}    W_{\mathrm{d}}=\int^{\tau}_{0} \mathrm{d}t \int_{-\infty}^{+\infty}\mathrm{d}x  \frac{\gamma J^{2}}{\rho} \label{Eq-overdamped-irr1}\end{equation}
with the probability flux
\begin{equation}    J \equiv - \frac{1}{\gamma} \left (  \frac{\partial U}{\partial x} \rho + \frac{1}{\beta} \frac{\partial \rho}{\partial x} \right ). \label{Eq-overdamped-flux}\end{equation}
Substituting the whole potential $U(x,t)=U_{0}(x,\lambda(t)) + U_{1}(x,t)$ and the instantaneous equilibrium distribution (\ref{Eq-overdamped-equilibrium}) into Eq.~(\ref{Eq-overdamped-irr1}), we obtain the special expression for the dissipated work (with the adaptation of shortcuts to isothermality):
\begin{equation}    W_{\mathrm{d}}=\int^{\tau}_{0} \mathrm{d}t \int_{-\infty}^{+\infty}\mathrm{d}x  \frac{\rho_{\mathrm{ieq}} }{\gamma} \left( \frac{\partial U_{1}}{\partial x}  \right)^{2} \ge 0 , \label{Eq-overdamped-irr2}\end{equation}
where equality holds only in quasi-static process. This implies that, in the overdamped situation, the dissipated work done in finite-time shortcuts to isothermality should be positive.
By using integration by part and considering Eq.~(\ref{Eq-overdamped-U1limitation}), we derive
\begin{eqnarray} W_{\mathrm{d}} && = \int^{\tau}_{0} \mathrm{d}t \int_{-\infty}^{+\infty}\mathrm{d}x \left(  \frac{\rho_{\mathrm{ieq}} }{\gamma} \frac{\partial U_{1}}{\partial x} \right)\frac{\partial U_{1}}{\partial x}  \nonumber \\ && =  -   \int^{\tau}_{0} \mathrm{d}t \int_{-\infty}^{+\infty}\mathrm{d}x  \beta U_{1} \left(  \frac{1}{\gamma \beta} \frac{\partial^{2} U_{1}}{\partial x^{2}} - \frac{1}{\gamma}\frac{\partial U_{0}}{\partial x} \frac{\partial U_{1}}{\partial x} \right) \rho_{\mathrm{ieq}} \nonumber \\ & & =  -   \int^{\tau}_{0} \mathrm{d}t \int_{-\infty}^{+\infty}\mathrm{d}x  \beta U_{1}  \left[ \left ( \frac{\mathrm{d} F}{\mathrm{d} \lambda} - \frac{\partial U_{0}}{\partial \lambda} \right ) \dot{\lambda} \right] \rho_{\mathrm{ieq}}  \nonumber \\ && =  -   \int^{\tau}_{0} \mathrm{d}t \int_{-\infty}^{+\infty}\mathrm{d}x  U_{1} \frac{\partial \rho_{\mathrm{ieq}}}{\partial t}  \nonumber \\ && =   \int^{\tau}_{0} \mathrm{d}t \int_{-\infty}^{+\infty}\mathrm{d}x  \rho_{\mathrm{ieq}} \frac{\partial U_{1} }{\partial t} ,     \label{Eq-overdamped-equivalence}\end{eqnarray}
which is equivalent to Eq.~(\ref{Eq-general-meanirrwork}). We have taken use of boundary conditions $\dot{\lambda}(0) = \dot{\lambda}(\tau) = 0$ at the last step of derivations.

\subsection{Underdamped Situation}

In the underdamped situation, one of the authors~\cite{Tu2014} has already given a proof of the non-negativity of the dissipated work for the Langevin dynamics. It was assumed previously that the potential is momentum independent. Here we generalize the derivation to the case that the potential depends on the momentum $p$ of the system. Using the definition of heat along a trajectory~\cite{Tu2014,Sekimoto2010,Seifert2012}, we may derive the mean heat absorbed from the medium:
\begin{eqnarray} Q = \int^{\tau}_{0} \mathrm{d}t \int_{-\infty}^{+\infty}\mathrm{d}x \int_{-\infty}^{+\infty}\mathrm{d}p (\mathbf{J} \cdot \nabla H),       \label{Eq-meanheat}\end{eqnarray}
where $\mathbf{J}$ represents the probability flux and $H$ is the total Hamiltonian. The gradient operator is defined as $\nabla \equiv \hat{\mathbf{x}} \partial / \partial x + \hat{\mathbf{p}} \partial / \partial p$, where $\hat{\mathbf{x}}$ and $\hat{\mathbf{p}}$ represent the unit vectors in the coordinate and the momentum of the system.
From the generalized Kramers equation (\ref{Eq-generalized-FP}) we have
\begin{equation}    \mathbf{J}   \equiv   \left(p + \frac{\partial U}{\partial p} \right)\rho \hat{\mathbf{x}} - \left( \frac{\partial U}{\partial x} + \gamma p + \gamma \frac{\partial U}{\partial p} +  \frac{\gamma}{ \beta \rho} \frac{\partial \rho}{\partial p}  \right)\rho \hat{\mathbf{p}}. \label{Eq-underdamped-flux}\end{equation}
Substituting Eq.~(\ref{Eq-underdamped-flux}) into Eq.~(\ref{Eq-meanheat}), we obtain
\begin{eqnarray} Q  =&&  -  \int^{\tau}_{0} \mathrm{d}t \int_{-\infty}^{+\infty}\mathrm{d}x \int_{-\infty}^{+\infty}\mathrm{d}p   \left[   \gamma \rho \left(  p + \frac{\partial U}{\partial p} \right) \right. \nonumber \\ && ~~~ \times \left.     \left(  p + \frac{\partial U}{\partial p} + \frac{1}{\beta \rho} \frac{\partial \rho}{\partial p} \right) \right].      \label{Eq-underdamped-heat}\end{eqnarray}
According to the definition of entropy along a trajectory~\cite{Seifert2005,Seifert2012}
\begin{equation}    s= - \mathrm{ln} \rho, \label{Eq-trajectory-entropy}\end{equation}
the rate of the mean entropy production may be defined as
\begin{equation}    \dot{S} \equiv \langle  \dot{s} \rangle = \left \langle - \frac{1}{\rho} \left( \frac{\partial \rho}{\partial t} + \frac{\partial \rho}{\partial x} \dot{x} + \frac{\partial \rho}{\partial p} \dot{p} \right)  \right \rangle. \label{Eq-entropy-rate1}\end{equation}
The ensemble average $\langle \cdots \rangle$ proceeds in two steps~\cite{Tu2014,Seifert2005}. First, we average over all trajectories which are located at given $x$ and $p$ at time $t$, leading to
\begin{equation}   \langle \dot{x} | x,p,t \rangle = \frac{J_{x}}{\rho}, ~~~ \langle \dot{p} | x,p,t \rangle = \frac{J_{p}}{\rho},  \label{Eq-local-velocity}\end{equation}
where $J_{x}$ and $J_{p}$ are the $x$ and $p$ components of $\mathbf{J}$ in Eq.~(\ref{Eq-underdamped-flux}), respectively.
Second, averaging over all $x$ and $p$ with the distribution $\rho(x,p,t)$ leads to
\begin{eqnarray} \dot{S} && = \left \langle - \frac{1}{\rho} \left( \frac{\partial \rho}{\partial t} + \frac{\partial \rho}{\partial x} \dot{x} + \frac{\partial \rho}{\partial p} \dot{p} \right)  \right \rangle \nonumber \\ && =  -   \int_{-\infty}^{+\infty}\mathrm{d}x \int_{-\infty}^{+\infty}\mathrm{d}p  \left( \frac{\partial \rho}{\partial t} + \frac{J_{x}}{\rho}\frac{\partial \rho}{\partial x} + \frac{J_{p}}{\rho}\frac{\partial \rho}{\partial p} \right) \nonumber \\ && = \int_{-\infty}^{+\infty}\mathrm{d}x \int_{-\infty}^{+\infty}\mathrm{d}p \rho \left[    \frac{\partial}{\partial x} \left( \frac{J_{x}}{\rho} \right)  + \frac{\partial}{\partial p} \left( \frac{J_{p}}{\rho} \right) \right] \nonumber \\ && =  -   \int_{-\infty}^{+\infty}\mathrm{d}x \int_{-\infty}^{+\infty}\mathrm{d}p \gamma \rho  \frac{\partial}{\partial p} \left(  p +  \frac{\partial U}{\partial p} + \frac{1}{\beta \rho} \frac{\partial \rho}{\partial p} \right) \nonumber \\ && =    \int_{-\infty}^{+\infty}\mathrm{d}x \int_{-\infty}^{+\infty}\mathrm{d}p  \gamma \frac{\partial \rho}{\partial p} \left(  p +  \frac{\partial U}{\partial p} + \frac{1}{\beta \rho} \frac{\partial \rho}{\partial p} \right) .     \label{Eq-entropy-rate2}\end{eqnarray}
Thus the entropy difference between the final state and the initial state follows
\begin{equation}    \Delta S =\int^{\tau}_{0} \mathrm{d}t \int_{-\infty}^{+\infty}\mathrm{d}x \int_{-\infty}^{+\infty}\mathrm{d}p  \gamma \frac{\partial \rho}{\partial p} \left(  p +  \frac{\partial U}{\partial p} + \frac{1}{\beta \rho} \frac{\partial \rho}{\partial p} \right). \label{Eq-entropy-production}\end{equation}
Combining Eqs.~(\ref{Eq-underdamped-heat}) and (\ref{Eq-entropy-production}), we finally achieve the mean dissipated work
\begin{equation} \begin{split}  W_{\mathrm{d}} & = T \Delta S - Q  \\  &=  \int^{\tau}_{0} \mathrm{d}t \int_{-\infty}^{+\infty}\mathrm{d}x \int_{-\infty}^{+\infty}\mathrm{d}p  \gamma \rho \left(  p +  \frac{\partial U}{\partial p} + \frac{1}{\beta \rho} \frac{\partial \rho}{\partial p} \right)^{2} \ge 0. \label{Eq-underdamped-irr1}\end{split}\end{equation}
Substituting the form of the whole potential $U(x,p,t) = U_{0}(x,\lambda(t))+U_{1}(x,p,t)$ and the instantaneous equilibrium distribution (\ref{Eq-underdamped-equilibrium}) into Eq.~(\ref{Eq-underdamped-irr1}), we derive the underdamped expression of the dissipated work (with the adaptation of shortcuts to isothermality):
\begin{equation}    W_{\mathrm{d}}=\int^{\tau}_{0} \mathrm{d}t \int_{-\infty}^{+\infty}\mathrm{d}x \int_{-\infty}^{+\infty}\mathrm{d}p \gamma\rho_{\mathrm{ieq}} \left( \frac{\partial U_{1}}{\partial p}  \right)^{2} \ge 0 . \label{Eq-underdamped-irr2}\end{equation}
This expression of the dissipated work implies similar property as the overdamped one. Since the auxiliary potential should be momentum dependent to realize the strategy of shortcuts to isothermality, the dissipated work in this situation is still determined to be positive. Again by using integration by part and considering Eq.~(\ref{Eq-underdamped-H1limitation}), we derive
\begin{eqnarray} W_{\mathrm{d}} && = \int^{\tau}_{0} \mathrm{d}t \int_{-\infty}^{+\infty}\mathrm{d}x \int_{-\infty}^{+\infty}\mathrm{d}p \left( \gamma\rho_{\mathrm{ieq}} \frac{\partial U_{1}}{\partial p}  \right)\frac{\partial U_{1}}{\partial p}  \nonumber \\ && =  -   \int^{\tau}_{0} \mathrm{d}t \int_{-\infty}^{+\infty}\mathrm{d}x \int_{-\infty}^{+\infty}\mathrm{d}p   \beta U_{1} \left(  \frac{\gamma}{\beta } \frac{\partial^{2}U_{1}}{\partial p^{2}} - \gamma p \frac{\partial U_{1}}{\partial p}  \right) \rho_{\mathrm{ieq}}  \nonumber \\ & & =   -   \int^{\tau}_{0} \mathrm{d}t \int_{-\infty}^{+\infty}\mathrm{d}x \int_{-\infty}^{+\infty}\mathrm{d}p   \beta U_{1} \left[ \left ( \frac{\mathrm{d} F}{\mathrm{d} \lambda} - \frac{\partial U_{0}}{\partial \lambda} \right ) \dot{\lambda} \right. \nonumber \\ & & ~~~~~  \left. + p \frac{\partial U_{1}}{\partial x}-\frac{\partial U_{0}}{\partial x}\frac{\partial U_{1}}{\partial p} \right] \rho_{\mathrm{ieq}} \nonumber \\ & & =   -   \int^{\tau}_{0} \mathrm{d}t \int_{-\infty}^{+\infty}\mathrm{d}x \int_{-\infty}^{+\infty}\mathrm{d}p \left(  U_{1} \frac{\partial \rho_{\mathrm{ieq}}}{\partial t} + \beta p U_{1} \frac{\partial U_{1}}{\partial x} \rho_{\mathrm{ieq}}  \right. \nonumber \\ & & ~~~~~  \left. - \beta  U_{1} \frac{\partial U_{0}}{\partial x}\frac{\partial U_{1}}{\partial p} \rho_{\mathrm{ieq}}   \right)     \nonumber \\ && =   -   \int^{\tau}_{0} \mathrm{d}t \int_{-\infty}^{+\infty}\mathrm{d}x \int_{-\infty}^{+\infty}\mathrm{d}p \left(  U_{1} \frac{\partial \rho_{\mathrm{ieq}}}{\partial t} - \frac{1}{2} \frac{\partial U_{1}^{2}}{\partial x} \frac{\partial \rho_{\mathrm{ieq}}}{\partial p}  \right. \nonumber \\ & & ~~~~~  \left. + \frac{1}{2} \frac{\partial U_{1}^{2}}{\partial p} \frac{\partial \rho_{\mathrm{ieq}}}{\partial x}   \right)     \nonumber \\ &&  =   -   \int^{\tau}_{0} \mathrm{d}t \int_{-\infty}^{+\infty}\mathrm{d}x \int_{-\infty}^{+\infty}\mathrm{d}p \left(  U_{1} \frac{\partial \rho_{\mathrm{ieq}}}{\partial t} - \frac{1}{2} \frac{\partial U_{1}^{2}}{\partial x} \frac{\partial \rho_{\mathrm{ieq}}}{\partial p}  \right. \nonumber \\ & & ~~~~~  \left. - \frac{1}{2} \frac{\partial^{2} U_{1}^{2} }{\partial x \partial p}  \rho_{\mathrm{ieq}}   \right)     \nonumber \\ &&   =   -   \int^{\tau}_{0} \mathrm{d}t \int_{-\infty}^{+\infty}\mathrm{d}x \int_{-\infty}^{+\infty}\mathrm{d}p \left(  U_{1} \frac{\partial \rho_{\mathrm{ieq}}}{\partial t} - \frac{1}{2} \frac{\partial U_{1}^{2}}{\partial x} \frac{\partial \rho_{\mathrm{ieq}}}{\partial p}  \right. \nonumber \\ & & ~~~~~  \left. + \frac{1}{2} \frac{\partial U_{1}^{2}}{\partial x} \frac{\partial \rho_{\mathrm{ieq}}}{\partial p}   \right)     \nonumber \\ &&  =  -   \int^{\tau}_{0} \mathrm{d}t \int_{-\infty}^{+\infty}\mathrm{d}x \int_{-\infty}^{+\infty}\mathrm{d}p U_{1} \frac{\partial \rho_{\mathrm{ieq}}}{\partial t}  \nonumber \\ && =   \int^{\tau}_{0} \mathrm{d}t \int_{-\infty}^{+\infty}\mathrm{d}x \int_{-\infty}^{+\infty}\mathrm{d}p \rho_{\mathrm{ieq}} \frac{\partial U_{1} }{\partial t} ,     \label{Eq-underdamped-equivalence}\end{eqnarray}
which is equivalent to Eq.~(\ref{Eq-general-meanirrwork}). We have also considered boundary conditions $\dot{\lambda}(0) = \dot{\lambda}(\tau) = 0$ at the last step of derivations.

\section{Time-reversal of shortcuts to isothermality\label{Sec-four}}

Here we will check the time-reversibility of shortcuts to isothermality, i.e., whether the time-reversal of shortcuts to isothermality can still drive the system from an equilibrium state to another one at the same temperature. In the general situation, the system Hamiltonian is given by Eq.~(\ref{Eq-total-Hamiltonian}), i.e.,
\begin{equation}H(\Gamma,t)=H_{0}(\Gamma,\lambda(t)) + U_{1}(\Gamma,t) .\label{Eq-timereversal-totalp1}\end{equation}
The time-reversed counterpart of the total potential can be derived as:
\begin{eqnarray} \bar{H}(\bar{\Gamma}, \bar{t}) && = H(\bar{\Gamma},t)|_{t=\tau-\bar{t}}  \nonumber \\ &&  =  H_{0}(\bar{\Gamma},\lambda(t))|_{t=\tau-\bar{t}} + [ \dot{\lambda}(t)f (\bar{\Gamma} , \lambda(t))] |_{t=\tau-\bar{t}} \nonumber  \\ &&  =  H_{0}(\bar{\Gamma},\bar{\lambda}(\bar{t})) -  U_{1}(\bar{\Gamma},\bar{t}),    \label{Eq-timereversal-totalp}\end{eqnarray}
where $\bar{H}$, $\bar{\Gamma}$, $\bar{\lambda}$, and $\bar{t}$ are the time-reversed counterparts of $H$, $\Gamma$, $\lambda$, and $t$. Here $\bar{\Gamma}$ is obtained from $\Gamma$ by reversing all the momenta. Hence, the time-reversal of the total system Hamiltonian equals the time-reversed counterpart of the original system Hamiltonian minus its corresponding auxiliary potential, which implies that the time-reversal strategy of shortcuts to isothermality can not achieve transitions between two equilibrium states. As a result, the subscript $``-"$ in Eq.~(\ref{Eq-third-property}) can not be omitted, which indicates us to estimate the free energy difference by using trajectories gathered from the time-reversal driving process.

As a paradigm, we consider overdamped motion of a Brownian particle in the potential
\begin{equation}U(X,\mathcal{T})=U_{0}(X,\Lambda(\mathcal{T})) + U_{1}(X,\mathcal{T}),\label{Eq-overdamped-poten}\end{equation}
where $X$, $\mathcal{T}$, and $\Lambda(\mathcal{T})$ represent the coordinate, time, and controlling parameter, respectively. $U_{1}(X,\mathcal{T})$ is the auxiliary potential of $U_{0}(X,\Lambda(\mathcal{T}))$. The motion is governed by the Langevin equation
\begin{equation}    \gamma \frac{\mathrm{d}X}{\mathrm{d} \mathcal{T}} = - \frac{\partial \left[ U_{0}(X,\Lambda(\mathcal{T})) + U_{1}(X,\mathcal{T}) \right]}{\partial X} +  \Xi(\mathcal{T}), \label{Eq-overdamped-LE1}\end{equation}
where $\Xi(\mathcal{T})$ represents Gaussian white noise.
According to Eq.~(\ref{Eq-timereversal-totalp}), the Langevin equation governing the time-reversal driving process follows
\begin{equation}    \gamma \frac{\mathrm{d} \bar{X}(\bar{\mathcal{T}})}{\mathrm{d} \bar{\mathcal{T}}} = - \frac{\partial \left( U_{0}(\bar{X}(\bar{\mathcal{T}}) ,\bar{\Lambda}(\bar{\mathcal{T}})) - U_{1}(\bar{X}(\bar{\mathcal{T}}),\bar{\mathcal{T}}) \right)}{\partial \bar{X}(\bar{\mathcal{T}})} +  \bar{\Xi}(\bar{\mathcal{T}}), \label{Eq-overdamped-LE2}\end{equation}
where $\bar{X}(\bar{\mathcal{T}}) = X (\tau - \bar{\mathcal{T}} )$ and $\bar{\Lambda}(\bar{\mathcal{T}}) = \Lambda(\tau-\bar{\mathcal{T}})$.
The free energy difference $ F(\Lambda(\tau)) - F(\Lambda(0))$ depends on the two endpoints of the controlling protocol, $\Lambda(0)$ and $\Lambda(\tau)$. In the time-reversal driving process, the free energy difference is $ F(\bar{\Lambda}(\tau)) - F(\bar{\Lambda}(0)) $.

Next, we introduce new variables, $t=\bar{\mathcal{T}}$, $\lambda(t) = \bar{\Lambda}(t)$, $x(t) = \bar{X}(t)$, and $\xi(t) = \bar{\Xi}(t)$. Then the free energy difference in the time-reversal driving process is transformed into $\Delta F = F(\lambda(\tau)) - F(\lambda(0)) $. Simultaneously, the Langevin equation (\ref{Eq-overdamped-LE2}) is transformed into
\begin{equation}    \gamma \frac{\mathrm{d} x(t)}{\mathrm{d} t} = - \frac{\partial \left( U_{0}(x,\lambda(t)) - U_{1}(x,t) \right)}{\partial x} +  \xi(t). \label{Eq-overdamped-LE3}\end{equation}
The above equation implies that we can forget the time-reversal process and interpret the ensemble average in Eq.~(\ref{Eq-third-property}) as the average over all trajectories evolving under the original system Hamiltonian minus its corresponding auxiliary potential.


\begin{thebibliography}{50}
\bibitem{auxiliaryH}Strictly speaking, $U_{1}(x,p,t)$ should be called auxiliary Hamiltonian, while for the sake of simplicity, we still name it auxiliary potential.
\end{thebibliography}
\end{document}